\RequirePackage[2020-02-02]{latexrelease}
\documentclass[reprint, superscriptaddress, aps, amsmath, amssymb, pre, onecolumn]{revtex4-2}
\usepackage{color}
\usepackage{dcolumn}
\usepackage{graphicx}
\usepackage{subfigure}
\usepackage{amsmath}
\usepackage{enumerate}
\usepackage{siunitx}
\DeclareSIUnit\Molar{M}
\DeclareSIUnit\cell{cell}

\usepackage{comment}
\usepackage{todonotes}

\begin{document}
	
	\title{Stochastic trade-offs and the emergence of diversification in \emph{E. coli} evolution experiments}

	\author{Roberto Corral L\'opez}
	\affiliation{Departamento de
		Electromagnetismo y F{\'\i}sica de la Materia and \\ Instituto Carlos I
		de F{\'\i}sica Te{\'o}rica y Computacional. Universidad de Granada.
		E-18071, Granada, Spain}
	\author{Samir Suweis}
	\affiliation{Dipartimento di Fisica e Astronomia Galileo Galilei, Universit{\`a} degli Studi di Padova, via Marzolo 8, 35131 Padova, Italy}
	\affiliation{Istituto Nazionale di Fisica Nucleare, 35131, Padova, Italy}

	\author{Sandro Azaele}
        \affiliation{Dipartimento di Fisica e Astronomia Galileo Galilei, Universit{\`a} degli Studi di Padova, via Marzolo 8, 35131 Padova, Italy}
	\affiliation{Istituto Nazionale di Fisica Nucleare, 35131, Padova, Italy}
	\affiliation{National Biodiversity Future Center,
		Piazza Marina 61, 90133 Palermo, Italy}
\affiliation{ Senior authors sharing the  leadership of the research}

	\author{Miguel A. Mu{\~n}oz}
	\affiliation{Departamento de
		Electromagnetismo y F{\'\i}sica de la Materia and \\ Instituto Carlos I
		de F{\'\i}sica Te{\'o}rica y Computacional. Universidad de Granada.
		E-18071, Granada, Spain}
             \affiliation{ Senior authors sharing the  leadership of the research}

\begin{abstract}
  Laboratory experiments with bacterial colonies under well-controlled conditions often lead to evolutionary diversification, where at least two ecotypes emerge from an initially monomorphic population. Empirical evidence suggests that such an "evolutionary branching" occurs stochastically, even under fixed and stable conditions. This stochasticity is characterized by: (i) occurrence of branching in a significant fraction, but not all, of experimental settings, (ii) emergence at widely varying times, and (iii) variable relative abundances of the resulting subpopulations across experiments.  Theoretical approaches to understanding evolutionary branching under these conditions have been previously developed within the (deterministic) framework of "adaptive dynamics."  Here, we advance the understanding of the stochastic nature of evolutionary outcomes by introducing the concept of "stochastic trade-offs" as opposed to "hard" ones.  The key idea is that the stochasticity of mutations occurs in a high-dimensional trait space and this translates into variability that is constrained to a flexible tradeoff curve in a lower-dimensional space. By incorporating this additional source of stochasticity, we are able to account for the observed empirical variability and make predictions regarding the likelihood of evolutionary branching under different experimental conditions. This approach effectively bridges the gap between theoretical predictions and empirical observations, providing insights into when and how evolutionary branching is more likely to occur in laboratory experiments.  \end{abstract}
	\maketitle

	\section{Introduction}

Understanding the origin and evolution of biological diversity is one of the central issues in evolutionary biology \cite{Darwin,Hutchinson,Gould}. Historically, diversification was primarily understood in the context of \emph{allopatry}, resulting from geographical isolation between subpopulations of a common ancestral species \cite{Muller, Mayr, Dobzhansky, Coyne, Turelli}. However, in recent years there has been a shift to the consideration of \emph{sympatric} diversification, which occurs without spatial segregation. This has been driven by mounting empirical evidence from both field and laboratory experiments \cite{Rainey, Friesen, Tyerman, Maclean, Saxer, Helling, Rosenzweig, Treves, Lenski, Linn, Barluenga, Savolainen, Ryan,Shnerb} as well as by the consolidation of adaptive dynamics theory as a robust theoretical  framework for addressing this type of eco-evolutionary problems \cite{Geritz1,Geritz2, Metz, Dieckmann96, Doebeli-Book, Dieckmann99,Geritz3, Ito, Doebeli-CF}.

    In the theory of adaptive dynamics, diversification is understood as the consequence of \emph{``evolutionary branching''} where two phenotypically distinct populations emerge from an ancestral monomorphic one (see \cite{Geritz1,Geritz2, Metz, Dieckmann96, Doebeli-Book} and below).  Adaptive dynamics assumes the presence of a ``resident population'', usually considered to be monomorphic, from which ``mutants'' of the dominant phenotype emerge at a small rate. This small mutation rate ensures that ecological and evolutionary processes occur on separate timescales.
These concepts are mathematically encapsulated in the notion of ``\emph{invasion fitness}'', which describes the dependence of a mutant's growth rate on the resident population. Specifically, the sign of this growth rate determines whether the mutant will go extinct or proliferate, eventually invading the resident population and becoming fixed. Thus, adaptive dynamics can be seen as the gradual phenotypic change of an entire population, evolving uphill in the fitness landscape \cite{Geritz1,Geritz2}. Importantly, owing to the frequency-dependent nature of the invasion fitness, the fitness landscape is a dynamic entity that changes as the population composition changes in phenotypic space.

Typically, this evolutionary process culminates in finding a (local) maximum of the fitness landscape, where the population stabilizes. However, sometimes it may result in the counterintuitive phenomenon of "evolutionary branching." This occurs when the endpoint of the gradient-ascending dynamics, which simultaneously modifies the fitness landscape itself, is a fitness \emph{minimum} rather than a maximum \cite{Geritz1,Geritz2, Metz, Dieckmann96, Doebeli-Book}. At this type of ``singular point'', the population can only increase its overall fitness by splitting into two sub-populations. Each sub-population can then separately ascend one side of the fitness landscape, resulting in diversification into two distinct lineages \cite{Geritz1,Geritz2,Doebeli-Book,Kisdi-review,Adaptive-speciation,Sireci}.

This form of adaptive diversification has been empirically observed in a myriad of experiments. For instance, isogenic and well-mixed populations of \emph{E. coli} have been reported to evolve into (at least) two coexisting ecotypes with different carbohydrate metabolisms, both in chemostats \cite{Helling, Rosenzweig, Treves} and in serial-dilution (batch) experiments \cite{Lenski, Friesen, Spencer, Spencer-2, Tyerman, LeGac, Herron}.
In chemostats, where the bacterial population is continuously fed solely with glucose, one emerging subpopulation specializes in consuming glucose, while the second specializes in scavenging acetate, a typical by-product of glucose metabolism. Conversely, in serial dilution experiments where both glucose and acetate are added in cycles of serial dilutions, the difference between the emerging phenotypes lies in the time lag required to switch to acetate consumption when glucose becomes depleted. For example, Treves et al. \cite{Treves} reported that 6 out of 12 independent replicate evolutionary experiments in a chemostat resulted in a stable polymorphism for up to 1750 generations. Similarly, Friesen et al. \cite{Friesen} observed clear dimorphism in 5 out of 12 independent replicate serial-dilution experiments after 1000 generations.

It is worth noting that despite the inherent stochasticity of mutations, evolutionary branching turns out to be quite reproducible, with a significant proportion of trials resulting in stable dimorphism \cite{Helling, Treves, Friesen, Saxer, Spencer-2}. Nevertheless, significant trial-to-trial variability is also observed; in particular: (i) diversification is reported in a large
fraction, but not all, of the experiments, (ii) it happens at broadly diverse times, and (iii) the population fraction of the two emerging ecotypes is variable across experiments \cite{Treves}.

Prior theoretical analyses within the adaptive dynamics framework have demonstrated that simplified eco-evolutionary models can undergo evolutionary branching under the described conditions \cite{Doebeli-CF, Friesen}. However, these approaches are deterministic in nature, leading to the prediction of branching occurring in all or none of the realizations based on certain modeling assumptions or parameter values. Specifically, to replicate branching, previous studies relied on the arbitrary selection of a mathematical function to represent a "trade-off" between the different metabolic choices to be optimized, while alternative choices of the trade-off curve did not result in branching \cite{Doebeli-CF, Friesen}. 
	
Trade-offs, often encapsulated by the idea that ``you cannot be good at everything,'' are fundamental in ecological and evolutionary modeling \cite{Fisher}. Typically, hard trade-offs, prevalent in modeling, reflect the notion that evolutionary pressures drive organisms toward a Pareto front \cite{ParetoSurfacesExp}. Here, improvements in one task come with a cost to efficiency in others \cite{Shoval}. In our context, a specific trade-off curve has been commonly assumed between the abilities to metabolize glucose and acetate. However, selecting a precise trade-off curve is often the least justified aspect of mathematical models. While conceptually well-founded, experimental measurement of trade-offs presents significant challenges \cite{Stearns, Garland1}. Furthermore, as extensively debated in the literature, the choice of a specific trade-off curve is pivotal as it fundamentally dictates the occurrence or absence of evolutionary branching \cite{Geritz2,Mazancourt}. An elegant solution to circumvent this conceptual challenge within the adaptive dynamics framework was proposed by de Mazancourt and Dieckmann \cite{Mazancourt}. They devised a strategy to assess, for any given eco-evolutionary model, the robustness of the potential for evolutionary branching as the mathematical form of the trade-off function is altered (see below).

Our main goal here is to provide a theoretical and computational framework to account for the observed variability in the outcomes of controlled evolutionary-branching experiments. We aim to achieve this by introducing the concept of  ``stochastic'' trade-offs, contrasting with the conventional "hard" trade-offs, within the adaptive dynamics framework. Here, stochastic trade-off implies that the population is not bound to evolve along a rigid, predefined curve (Pareto front) in phenotypic space, as is the case with hard trade-offs. Instead, it operates within a more flexible and extended region, allowing for greater exploration of phenotypic diversity. As we will show, by adjusting the width of this region, which reflects the flexibility of the underlying bacterial metabolic constraints, it is possible to effectively regulate the degree of variability.

Let us remark that very recent theoretical works have also reported variability in the emergence of diversification across replicates \cite{Amicone,Good}. In particular, in \cite{Amicone} the strict trade-off is replaced by an upper bound for the consumption rates of the different resources (i.e. the metabolic traits) and the authors investigate the effect of clonal interference (i.e. strong-mutations \cite{Desai}) and its role in fostering diversification under varying mutation rates. Conversely, in \cite{Good}, the authors examine the interplay between selection and diversification by controlling the relative rates of purely fitness mutations and resource-strategy mutations, exploring the resulting regimes. Thus, in both studies, variability in diversification is controlled by mutation rates. We also find that larger mutation rates (beyond the infinitesimally small mutation limit) enhance variability. However, in contrast to previous studies, we propose an alternative source of variability that is independent of changes in the mutation rate: a stochastic trade-off model, where variability arises from the presence of a high-dimensional trade-off manifold representing the underlying metabolic constraints between all evolution-susceptible variables. Specifically, in our approach the variability of metabolic traits in the high-dimensional manifold is captured by a stochastic evolution of phenotypic traits in a lower dimensional space (see below).

Given the limited availability of experimental data, it is challenging to determine which mechanism accounts best for the observed variability, and it is also possible that multiple factors may contribute. However, we believe that the  framework we introduce is especially well-suited for investigating this problem because: (i) we utilize a biologically realistic ecological model fed with empirically fitted data to quantitatively describe the chemostat experiments \cite{Helling, Rosenzweig, Treves} (although the concepts could also be extended to analyze serial-dilution experiments); (ii) we propose a straightforward extension of the well-known framework of adaptive dynamics by adding stochasticity into the trade-off which allows for a controlled inclusion of variability; (iii) we generalize our results for a large family of trade-off curves, and (iv) we formulate additional theoretical predictions about the emergence of diversification under varied experimental scenarios.

The paper is structured as follows: Section \ref{sec:Model} introduces the eco-evolutionary model and defines the concept of stochastic trade-offs. Section \ref{sec:Results} outlines the main findings, while Section \ref{sec:Discussion} offers conclusions and delineates future research avenues.

\section{Eco-evolutionary model building} \label{sec:Model}
	
\subsection{Ecological dynamics}

To gain insight into the results of \emph{E. coli} evolution experiments in chemostats, we first construct a simple yet biologically realistic ecological model, inspired by existing modeling approaches \cite{Doebeli-CF, Gudelj, Liao}. This model aims to quantitatively describe the bacterial population and integrate key aspects of the experimental setups \cite{Helling, Rosenzweig, Treves}. Specifically, it encompasses the following components and characteristics:

	\begin{itemize}
		\item Initially, the bacterial population is monomorphic, consisting of a large number  of phenotypically identical individuals (of the order of $\SI{e13}{\cell\per \liter}$ \cite{Gudelj}), and it is cultured in a glucose-limited chemostat \cite{Rosenzweig, Treves}.
		
		\item 
Under the conditions of the experiment, when bacteria metabolize glucose, they produce acetate due to overflow metabolism (a phenomenon where fermentation is preferred over respiration). This results in the excretion of acetate in proportion to the amount of glucose consumed \cite{Hwa, Rosenzweig}.
			
		\item 	
The excreted acetate acidifies the medium which leads to an overall slowing down of  the total population growth rate \cite{Gudelj}.
		
\item Even if a strain emerges with the ability to metabolize acetate, its consumption is typically repressed in the presence of glucose, which is the preferred resource \cite{Liao}.
	\end{itemize}

        Considering these ingredients one can write the following set of (deterministic) differential equations for the total number of cells, $N(t)$, and the concentrations of glucose, $G(t)$, and acetate, $A(t)$, respectively, that represent the ecological (not evolutionary) part of the dynamics:
	\begin{widetext}
	\begin{eqnarray}
	\frac{dN}{dt}&=&\gamma\left(V_{g}\frac{G}{K_{g}+G}+V_{ a}\frac{A}{K_{a}+A}\frac{C_{g}}{C_{g}+G}\right)\frac{C_{a}}{C_{a}+A}N-DN \label{eq:Model1}\\
		\frac{dG}{dt}&=&D (G_0-G) -V_{g}\frac{G}{K_{g}+G}N \label{eq:Model2} \\
	\frac{dA}{dt}&=&\beta V_{g}\frac{G}{K_{g}+G}N-V_a\frac{A}{K_{a}+A}\frac{C_{g}}{C_{g}+G}N-DA. \label{eq:Model3}
	\end{eqnarray}	
	\end{widetext}
Observe first that all growth kinetics are described using Monod's functions \cite{Monod}. For example, $V_{g}\frac{G}{K_{g}+G}$ (resp. $V_{a}\frac{A}{K_{a}+A}$) represents the consumption  of glucose (resp. acetate) at a maximum growth rate per capita 
\(V_{g}\)  (resp. \(V_{a}\))  and half-saturation constant \(K_{g}\) (resp. \(K_{a}\)). The factor involving \(C_g\) controls the repression of acetate consumption due to the presence of glucose, while the one with \(C_a\) represents the inhibition of overall growth caused by acetate-induced acidification of the medium. \(\gamma\) is the conversion constant for biomass production from resources,
\(\beta\) determines the fraction of consumed glucose excreted as acetate, \(D\) is the chemostat's dilution rate, and \(G_0\) is the rate at which glucose is supplied to the chemostat. The corresponding numerical values of all these parameters are taken from experimental measurements \cite{Helling, Rosenzweig, Treves} and are specified in SM Table I. 

Thus, in a nutshell, the population \(N\) increases due to the consumption of glucose and/or acetate. However, consumption of acetate is inhibited due to catabolite repression and overall growth is limited due to the acidification of the medium caused by acetate.

\subsection{Evolutionary dynamics}

To complete the eco-evolutionary model, we must define the evolutionary part of the dynamics.
 This involves determining which ecological parameters (e.g., constants in Eqs. \ref{eq:Model1}-\ref{eq:Model3}) are subject to change over time and which remain constant throughout the evolutionary process.
Based on experimental findings \cite{Helling, Rosenzweig, Treves}, it is evident that the abilities to metabolize both glucose and acetate evolve over the course of the experiments. Therefore, parameters such as $V_{a,g}$, $K_{a,g}$, and $C_g$ are all potential candidates for evolution. For the sake of simplicity, we opt for $V_g$ and $V_a$ ---the maximum growth rates in glucose and acetate, respectively, represented collectively as $\textbf{V}=(V_g,V_a)$---  to serve as evolving parameters. These parameters are allowed to vary across generations, defining the
\emph{metabolic strategy} of each strain.  

To incorporate evolution using adaptive dynamics, it is assumed that once the ecological dynamics reach a steady state (i.e., a state where Eqs. \ref{eq:Model1}-\ref{eq:Model3}  with $\textbf{V}=(V_g, V_a)$ stabilize), a "mutation" occurs. This implies that ecological and evolutionary processes are considered to occur at well-separated time scales.
This mutation generates an individual mutant with a slightly different phenotype $\textbf{V}'=(V'_g, V'_a) \neq (V_g, V_a)$. The invasion fitness of this mutant, defined as its per-capita growth rate within the existing resident population, can be expressed (from Eq. \ref{eq:Model1}) as:
	\begin{widetext}
	\begin{equation}
		f(\textbf{V'},\textbf{V})=\gamma\left(V'_{g}\frac{G^*}{K_{g}+G^*}+V'_{ a}\frac{A^*}{K_{a}+A^*}\frac{C_{g}}{C_{g}+G^*}\right)\frac{C_{a}}{C_{a}+A^*}-D
	\end{equation}
	\end{widetext}
where $G^*$ and $A^*$ represent the steady-state values of the glucose and acetate concentrations, respectively, which depend on the values of both $V_g$ and $V_a$ of the resident population.

Using the formalism of adaptive dynamics, in the (deterministic) limit of small mutations, the evolution in phenotypic space is determined by the \emph{canonical equation}, describing how the average phenotype in the population climbs up the fitness landscape \cite{Dieckmann96, Doebeli-Book, Ito}:
	\begin{equation}\label{eq:canonical1}
	\frac{d\textbf{V}}{dt}= \mu \nabla_\textbf{V'} f(\textbf{V'}, \textbf{V})\Bigr|_{\textbf{V'}=\textbf{V}}
	\end{equation}
where  $\nabla_\textbf{V'} =[\frac{\partial}{\partial{V_1'}}, \frac{\partial}{\partial{V_2'}}]$ is the fitness gradient and $\mu$ is the mutation rate, which in the  simplest case adopted here is just a constant.

	At this point, a specific trade-off function, denoted as \(V_a = h(V_g)\), is typically introduced. This function reflects the biological fact that enhancing one metabolic strategy comes at the expense of efficiency in the consumption of the other, so that both strategies cannot be altered independently \cite{Fisher}. This implies that the canonical equation (Eq.(\ref{eq:canonical1})) needs to be modified by an additional term to account for such a hard constraint as specified in  Supplementary Material (SM) Eq.(1).
Analyzing the fixed points of the resulting equation, along with the properties (first and second derivatives) of the modified fitness gradient at these points, one could determine whether the course of evolution (in the deterministic limit) leads to an evolutionarily stable population, or to the emergence of evolutionary branching \footnote{Analytical solutions to this problem are often challenging, so one frequently resort to pairwise invasibility plots for insights (see \cite{Geritz2, PIP} for further details).}.

As discussed in the introduction and explicitly demonstrated in the Results section, the previously-described adaptive-dynamics model coupled with a hard trade-off is deterministic. This means that it either exhibits branching for every model realization or it does not for any, depending on the specific shape of the considered trade-off curve. Consequently, it fails to elucidate the observed variability in the experiments such as the timing of diversification (branching) events and the proportion of the two emerging ecotypes over time.
Let us remark that stochastic effects in the context of adaptive dynamics have already been considered in the literature. Wakano and Iwasa \cite{Wakano} (and more recently some of us \cite{Sireci}) have shown that finite-size effects in the population number may prevent the system from undergoing evolutionary branching, even though it is theoretically predicted to emerge at a deterministic level.  However, these effects become negligible for the very large population sizes under consideration (here, $N\sim 10^{13}$).

To address stochasticity, we adopt a different approach. As stated in the introduction we
are interested in explaining properties of the evolutionary experiments such as the time at which a diversification (branching) event occurs and the fraction of the two emerging ecotypes at a certain time. Therefore, we need to build a framework capable of reproducing actual stochastic evolutionary trajectories that follow, on average, the behavior described by the canonical equation, but also allow for variability.

Ideally, one would aim to construct a bottom-up approach by defining an individual-based model where each individual has its own associated phenotype, allowing for evolution through the combined processes of mutation and selection to occur \cite{Sireci, Dieckmann96}. However, given the size of empirical populations, typically around  $N \sim 10^{13}$, such a "microscopic" description is currently computationally unfeasible. Therefore, as an alternative, we employ a framework adapted from \cite{Caetano} (see SM Sec. III), which represents an effective "mesoscopic" description of the evolutionary trajectories. This approach does not directly deal with individual organisms but rather with "groups" of them (subpopulations) that share the same phenotype, i.e. individuals are clustered in groups and one writes dynamical equations for the abundances of such groups. Essentially, this involves considering a collection of possible coexisting subpopulations, which necessitates (i) expanding the set of ecological equations to account for multiple subpopulations instead of just one,  (ii) running the ecological dynamics for a sufficiently large time as to reach a steady state, (iii) introducing an additional equation for a new mutant subpopulation whose phenotype slightly differs from the subpopulation from which it  derives, and (iv) iterating this eco-evolutionary process.

Two remarks are in order before proceeding: First, let us recall that this framework, where the ecological dynamics reach stationarity before new mutations appear ---as customarily assumed in adaptive dynamics \cite{Doebeli-Book}--- corresponds to the strong-selection weak-mutation regime \cite{Desai}. Note that for very large bacterial populations such as those studied here, the number of mutations per unit time in the entire population might be high \cite{Desai}. Thus, the key implicit assumption is that, even if generic mutations occur at a high rate, those directly affecting pathways involving the metabolism of acetate and glucose are much rarer, allowing us to assume a large separation of timescales between the ecological and evolutionary dynamics. Experimental findings support this assumption, showing that all acetate-scavenging strains acquired their ability to metabolize acetate through mutations in a specific gene \cite{Treves}. Therefore, although the overall mutation rate might be high, mutations impacting that particular gene are much less frequent.

Second, to achieve the strict limit of perfect time-scale separation in computational analyses, one would need to run the ecological dynamics for an arbitrarily long time to ensure that a steady state has been reached. Since the simulation time is necessarily finite, what one actually analyzes is a limit of a small, but not infinitely small, mutation rate, \i.e., a weak (but not infinitely-weak) mutation limit.  This is why the algorithm accommodates a set of possible subpopulations, i.e., a distribution in phenotypic space, instead of a single population. In any case, one can perform analyses by progressively expanding the time of the ecological dynamics to verify that the resulting distributions become sharper as the time scale is enlarged.

\subsection{Stochastic trade-offs}

\begin{figure*}
	\includegraphics[width=0.95\textwidth]{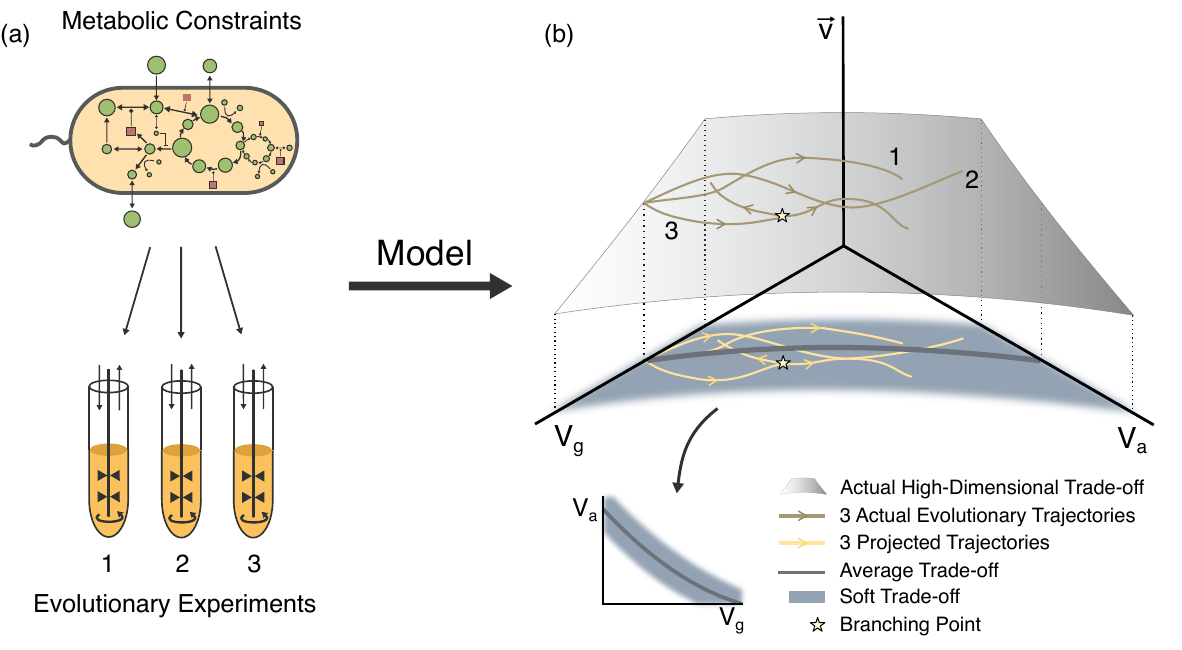}
	\caption{\textbf{Schematic representation of the concept of stochastic trade-off.} a) Metabolic constraints of bacteria influence the outcome of evolutionary experiments.  b) These constraints can be represented by a (topologically complex) manifold (gray shadowed surface), in the multidimensional space formed by all the relevant variables for evolution, which in our case are $V_g$, $V_a$ and the rest, $v$. Starting from the same point, the different evolutionary trajectories corresponding to different realizations of the experiments can move freely in the high-dimensional constraint manifold (golden curves), possibly giving rise to diversification events if they reach a singular point (yellow star). Thus, the trajectories projected in the ($V_g,V_a$) plane (yellow lines) are restricted to a certain zone (blue area), but can be different from each other. We model this constraint region, that is, the stochastic trade-off in the lower-dimensional manifold, by adding a temporarily-correlated noise on top of a certain fixed curve that stands as the average of possible trajectories (inset plot).}
	\label{fig:MetabolismTradeoff}
\end{figure*}

We here introduce an additional source of stochasticity to account for the experimental variability. Specifically, we suggest the concept of a stochastic trade-off, wherein the population is not confined to a fixed trade-off curve in phenotypic space but can instead wander within a broader region with fuzzy boundaries \cite{Roff}.  

The concept of stochastic tradeoff can be rationalized as follows (see also Fig.\ref{fig:MetabolismTradeoff}).	
The underlying metabolic constraints of bacteria are indeed the same for every realization of the evolutionary experiments. However, such constraints involve complex relations between the many variables that are susceptible to evolution. In our case, we explicitly consider the evolution of $V_g$, $V_a$ but, in principle, there are many other variables, denoted generically as $\vec{v}$, that could be affected by overall (energetic/metabolic/...) constraints. These intricate relationships can be mathematically modeled as a multidimensional constraint manifold, akin to a high-dimensional ``Pareto front'' as depicted in Fig. \ref{fig:MetabolismTradeoff} \cite{ParetoSurfacesExp, PlasticTradeoff}. In other words, the actual trade-off involves a high-dimensional manifold, rather than the two-dimensional one characterizing metabolic preferences.

In such a framework, different experimental realizations,  originating from the same point in the high-dimensional manifold (or Pareto front), can produce diverse trajectories that diverge along distinct paths inside the manifold due to the emergence of different mutations, which are intrinsically stochastic. Therefore, as illustrated in Fig.\ref{fig:MetabolismTradeoff}, the evolutionary trajectories when projected  in the two-dimensional $(V_g,V_a)$-plane are generally not confined to a one-dimensional curve and they may vary across experimental realizations. It is thus reasonable to consider the effect of metabolic constraints as defining a wide region in the lower-dimensional space of metabolic preferences
rather than as a strict (one-dimensional) curve. 
		
More specifically, we describe these constraints in the $(V_g,V_a)$-plane as a stochastic process (modelled as an Ornstein-Uhlenbeck process with amplitude $\sigma$ and temporal correlation $\tau$, as explicitly shown in SM Sec. II), added as an orthogonal perturbation to some central trade-off curve. This curve stands as the average of trajectories and can be chosen with large flexibility without affecting the conclusions (see SM Sec. II). Depending on the amplitude and temporal correlations of the additional stochasticity, trajectories are constrained to wander at tunable distances of the central trade-off curve, allowing us to model the flexibility of the metabolic constraints in the $(V_g,V_a)$-plane and in turn the variability of the trajectories. 

It is important to note that, although selection in the two-dimensional phenotypic plane tries to push trajectories away from the average curve in the direction of increasing fitness, where both $V_g$ and $V_a$ growth, our model "forces" the trajectories to stay close to the average curve. This serves as an effective way to model the constrains arising from the higher-dimensional manifold where mutations can have a positive, negative or neutral effect when projected onto the $(V_g-V_a)$-plane.

In summary, the stochastic trade-off allows for the population to navigate and randomly explore a broad phenotypic region (in the $(V_g,V_a)$-plane), effectively capturing the impact of metabolic constraints, rather than being restricted to a fixed trade-off curve.
 
	\section{Results} \label{sec:Results}
	\subsection{Reproducibility and variability of experimental results}\label{sec:Variability}
	
	\begin{figure*}
		\includegraphics[width=1.0\textwidth]{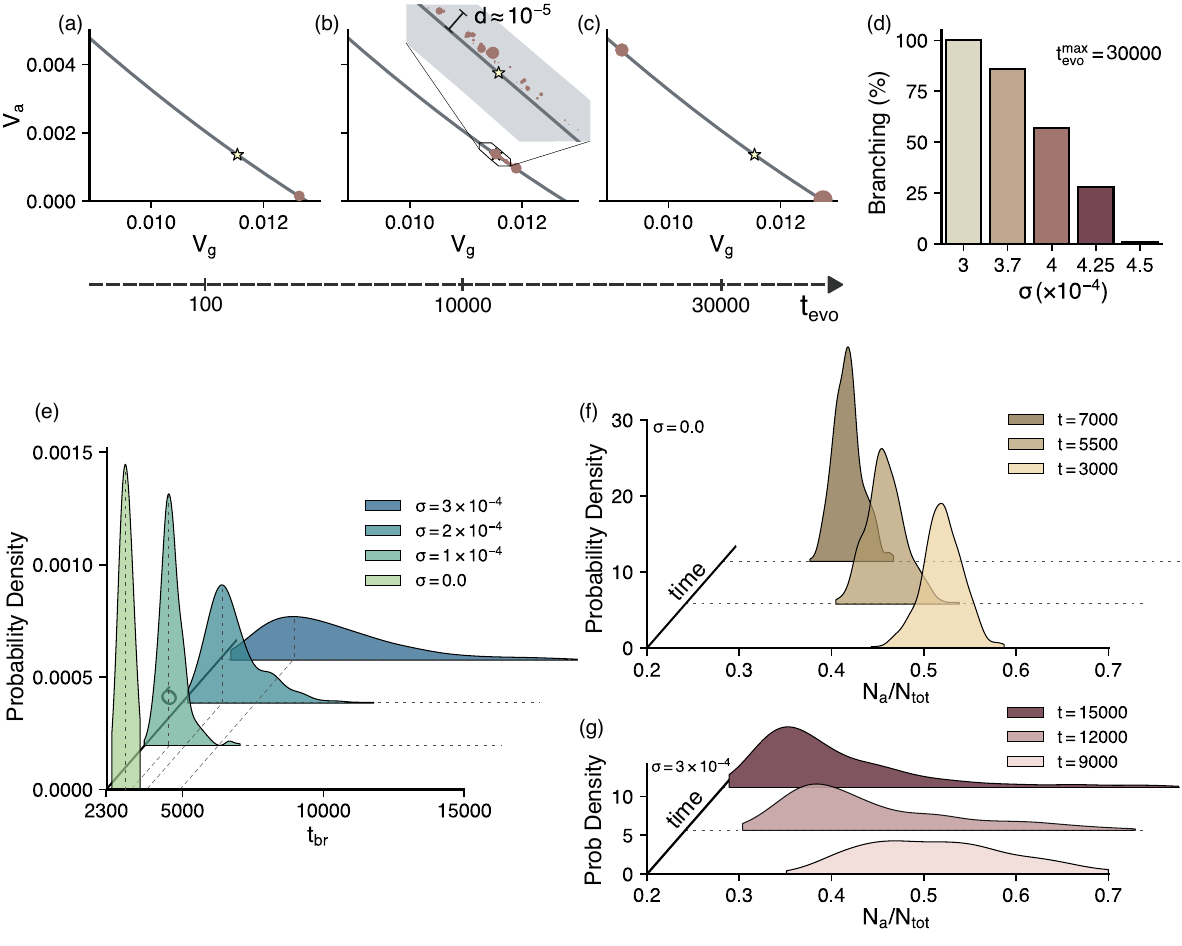}
		\caption{\textbf{Reproducibility and variability of evolutionary branching in our mesoscopic model with a stochastic trade-off.}  (a-c) Sub-populations in phenotypic space (represented by a red circle whose radius is proportional to its abundance) for three different times along the evolutionary trajectory, pre-branching (a), during branching (b) and after branching (c), for one specific realization with $\sigma=4\cdot10^{-4}$. The inset in b) is a zoom in the region close to the branching point (yellow star), where $d\approx10^{-5}$ is the maximum distance from a subpopulation to the average of the stochastic trade-off. (d) Percentage of realizations (number of them $N_r=200$) that undergo evolutionary branching for different amplitudes of the stochastic trade-off. (e-g) Probability density distributions across experiments (number of realizations $N_{r}=500$) of: branching times for different stochastic trade-off amplitudes (with dashed lines highlighting the shift in the peaks of the distributions) (e); and fraction of acetate scavengers at different times for $\sigma=0$ (f) and $\sigma=3\cdot10^{-4}$ (g) (see SM Sec. IV). System's parameters for all the plots are: $\beta=0.25$ and those specified in SM Table I where the shown $V_{g/a}$ are rescaled with respect to the table value as $V_{a/g}=\gamma V_{a/g}^\text{table}$ with $\gamma=4.5\cdot10^{10}$. Stochastic trade-off parameters are: $(a_{tr}, b_{tr}, c_{tr}, \tau)=(83.3,-3.5,0.03,10)$ (see SM Sec. V for definitions and further information). Evolutionary simulations are carried out as described in SM Sec. III.}
		\label{fig:Result1}
	\end{figure*}
	
	As illustrated in Fig.\ref{fig:Result1}, computational analyses of the eco-evolutionary (mesoscopic) model coupled with a stochastic trade-off (as defined in SM Sec. II) reveal that it is able to account for evolutionary branching as well as for the variability evinced in the experiments \cite{Treves}.  In particular, Fig.\ref{fig:Result1}(a-c) shows three temporal snapshots of the evolution of the population in phenotypic space for a specific realization of the model with the stochastic trade-off (as illustrated in the inset of panel b) that ended up diversifying in two highly-specialized sub-populations.
	
	Panel (a) represents the initial phase of the evolutionary picture, in which a glucose-specialist population that produces acetate evolves its ability to assimilate acetate because it is evolutionarily-favoured, so that it starts moving progressively ``uphill''.  Panel (b) shows the phase in which the population has accumulated some variability and --upon arriving to a singular point--- splits in two main ecotypes due to disruptive selection: one more specialized in acetate and the other still relying predominantly on glucose consumption.
Panel (c) illustrates the final phase, where the two branched types diverge, each becoming increasingly specialized in metabolizing one of the two carbon sources. This divergence thus gives rise to the two distinct ecotypes observed in the experiments: the glucose specialist and the acetate scavenger.

 This evolutionary picture is fully compatible with the one described in the experiments \cite{Rosenzweig}.
	 
On the other hand, Fig.\ref{fig:Result1}d summarizes the amount of realizations that, starting from a glucose-specialist population and with the same parameters as in panel (a), ended up undergoing evolutionary branching at a maximum evolutionary time (arbitrarily fixed to  $t_\text{evo}^\text{max}=3\cdot10^4$) for different amplitudes of the stochastic trade-off region, as obtained by varying the noise parameter, $\sigma$, for constant $\tau$. In this way, by tuning the noise amplitude ---or equivalently the flexibility of the metabolic constraints--- the model is able to account for the experimental fact that diversification is observed in variable proportion of the experiments. 

Furthermore, as shown in panels (e-g) the model can also explain the empirically-observed variability in  both  the branching times and the proportion of acetate scavengers after population splitting. Indeed, panel (e) reveals that branching times ---estimated as explained in SM Sec. IV--- are variable and their distribution depends on the noise amplitude, $\sigma$. One can observe that, in the case with a "hard" trade-off (i.e. $\sigma=0$), branching occurs almost always at the same time (strongly peaked distribution), but, 
the standard deviation of the noise $\sigma$ ---or equivalently, the amplitude of the trade-off--- increases, the distributions of branching times become wider and shift further to the right, corresponding to longer branching times. On the other hand,  Fig.\ref{fig:Result1}f shows that the distribution of the fraction of acetate scavengers at different times after branching is sharply peaked in the case with a hard trade-off, whereas the case of stochastic trade-off  in Fig.\ref{fig:Result1}g shows a much broader distribution.

As previously mentioned, the finite widths that emerge in the absence of stochasticity in the trade-off (i.e., for $\sigma=0$) arise from the limited time allowed for the ecological dynamics to stabilize. This is because we assume weak mutations, but not the infinitely-weak mutation limit. As a result, clonal interference and the intrinsic randomness of the evolutionary algorithm may occur (see SM Sec. III). We have computationally verified that the distributions gradually narrow as the time allowed for the ecological dynamics to reach equilibrium is extended (data not shown).

	Therefore, assuming the data could be explained with a hard trade-off, one should see practically the same fraction for different realizations if measured at the same times, whilst assuming a stochastic trade-off, one should observe significant differences even if measured at same times. Although experimental data for branching times and fractions is scarce to allow for a quantitative comparison, our results suggest that the empirical observations are best represented by the stochastic case \cite{Treves}.

\subsection{Branching feasibility and trade-off (in)dependence}

\begin{figure*}
	\includegraphics[width=1.0\textwidth]{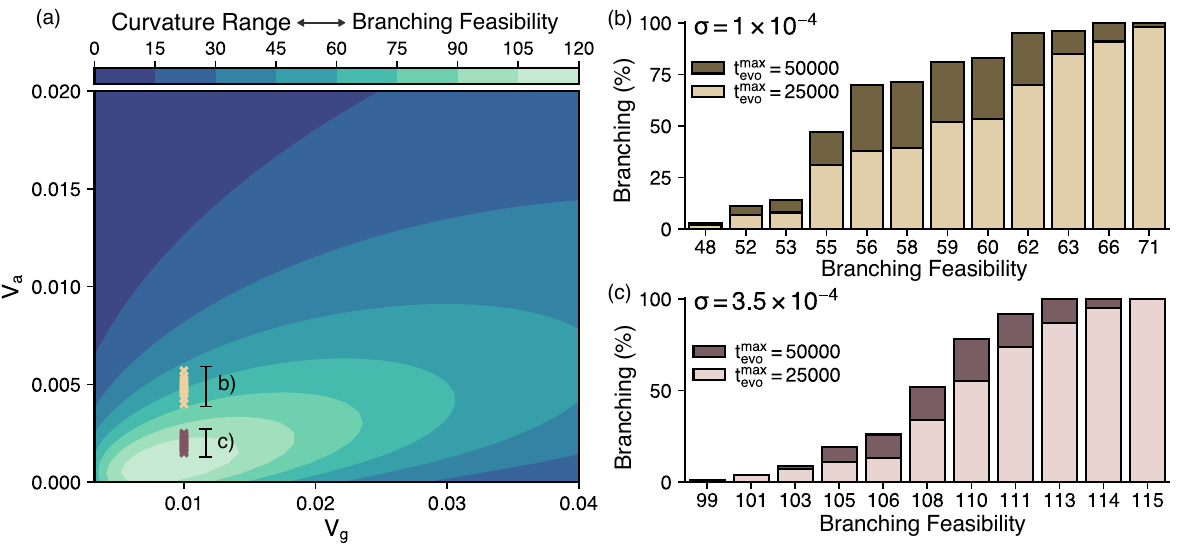}
	\caption{\textbf{Branching feasibility and the fraction of realizations that undergo branching.} 
		a) Branching feasibility plot: contour plot that illustrates in phenotypic space the range of local curvatures of the hard trade-off that allow evolutionary branching. This curvature range can also be interpreted (with the stochastic trade-off idea) as the likeliness of having evolutionary branching at each point.
		b-c) Percentage of realizations (total number, $N_r=200$) of the eco-evolutionary model with the stochastic trade-off that undergo evolutionary branching for two different maximum evolutionary times $t_\text{evo}^\text{max}=2.5\cdot10^4$ and $t_\text{evo}^\text{max}=5\cdot10^4$ at different points in phenotypic space (that have different curvature range or equivalently different branching feasibility). Points in (b) are those in yellow in panel (a) and $\sigma=10^{-4}$. Points in (c) are those in bourdeaux in panel (a) and $\sigma=3.5\cdot10^{-4}$. System's parameters are: $\beta=0.25$ and those specified in SM Table I where the shown $V_{g/a}$ are rescaled with respect to the table value as $V_{a/g}=\gamma V_{a/g}^\text{table}$ with $\gamma=4.5\cdot10^{10}$. Stochastic trade-off parameters are: $\tau=10$ and the others are obtained for each point shown in panel a) as specified in SM Sec. II with $\delta=0.25$. Eco-evolutionary simulations are carried out as described in SM Sec. III.}
	\label{fig:Result2}
\end{figure*}

	As already discussed in the introduction, the specific evolutionary outcomes of eco-evolutionary models depend crucially on the form of the trade-off \cite{Geritz2, Mazancourt}. Therefore, in order to generalize the results found in Sec.\ref{sec:Variability} without relying on a specific form of the (mean) trade-off curve, one needs a framework that is able to discern whether a given ecological model equipped with a certain set of parameter values, is able to undergo evolutionary branching (or not) independently of the specific form of the trade-off.
	
        To this aim, we make use of the formalism proposed by de Mazancourt and Dieckmann \cite{Mazancourt}. As originally developed, such a framework allows one to analyse geometrically the possible evolutionary outcomes that can occur in the deterministic setting without imposing a specific mathematical function for the trade-off (for the sake of completeness, an overview of the method is described in SM Sec. I).  In particular, within this formalism one can construct plots such as that of Fig. \ref{fig:Result2}a, providing information on how independent is evolutionary branching on the specific mathematical form of the hard trade-off.
	More specifically, it gives at each point in phenotypic space, a measure of the range of possible local curvatures of the hard trade-off function that allows the model to undergo evolutionary branching at such a point, thus quantifying the robustness of branching against changes in the shape of the trade-off function (see SM Sec. I). Regions where the range of possible local curvatures is small (deep blue in Fig. \ref{fig:Result2}a), allow the model to undergo evolutionary branching deterministically only for very specific (fine-tuned) trade-off curves. Instead, in regions where the curvature range is high (light green in Fig. \ref{fig:Result2}a), evolutionary branching occurs deterministically for a larger family of possible trade-off functions (see SM Sec. I for further details).
	
	Now, we would like to see how this (deterministic) formalism could be reinterpreted when we include the idea of stochastic trade-offs. It turns out that using stochastic trade-offs, the previous plots (e.g. Fig.\ref{fig:Result2}a) can be interpreted as ``\emph{evolutionary-branching feasibility plots}'', i.e. graphs showing the regions in phenotypic space where it is more likely to see an evolutionary branching event. In other words, regions in phenotypic space where branching is more robust against changes in the specific form of the hard trade-off correspond to regions in phenotypic space where branching is more likely to happen with the stochastic trade-off and vice-versa. 
	
	To justify this claim, we employ without loss of generality our eco-evolutionary model with a certain stochastic trade-off, whose average curve produces branching deterministically (i.e. in the $\sigma=0$ limit) at a selected point in phenotypic space (see SM Sec. II for further details on how the stochastic trade-off is built). Then, as illustrated in Fig.\ref{fig:Result2}b-c, points where deterministic branching depends weakly on the specific form of the hard trade-off ---i.e. points where the range of local curvatures of the hard trade-off that allow branching is larger (brighter colours in Fig.\ref{fig:Result2}a)--- correspond in the stochastic trade-off case to more realizations that end up experiencing diversification at that point and vice-versa.  Note, however, that in this framework we can only work out the relative probabilities of branching. In other words, we can only indicate whether or not one branching condition is relatively more likely compared to another one.

	Let us however remark, that this mapping between the range of curvatures and feasibility of branching is quantitatively conditioned by the maximum evolutionary time of the realizations, $t_\text{evo}^\text{max}$. As both panels (b) and (c) show, the higher $t_\text{evo}^\text{max}$, the larger is the probability to observe evolutionary branching. This intuitive fact
 ---i.e. the increase in the number of experiments that show diversification with the duration of such experiments--- is indeed observed also experimentally \cite{Treves}. Furthermore, Fig.\ref{fig:Result2}b-c illustrates that the mapping is also quantitatively affected by the amplitude of the stochastic trade-off (controlled by $\sigma$). In agreement with Fig.\ref{fig:Result1}d, the smaller $\sigma$, the higher the probability of having evolutionary branching. In fact, in the hard trade-off limit where $\sigma=0$ and the trade-off reduces to its average curve (which is chosen such that it undergoes branching), evolutionary branching occurs deterministically at almost the same times (as shown in Fig.\ref{fig:Result1}g), independently of the point in phenotypic space.

	Therefore, interpreting Fig.\ref{fig:Result2}a as a branching-feasibility plot, we observe that branching is more likely for low values of $V_a$, which agrees with the reported picture of the evolution experiments given in \cite{Rosenzweig}. As shown in Sec.\ref{sec:DifferentConditions} of the SM, the specific contour pattern depends highly on the conditions of the experiment, that is, the concentration of glucose in the environment supplying the chemostat, $G_0$, the dilution rate, $d$, and the proportion of the consumed glucose that gets excreted as acetate, $\beta$. Thus, Fig.\ref{fig:Result2}a reveals essentially that the system, under these specific environmental conditions, is rather likely.

	\subsection{Theoretical predictions for different experimental values} \label{sec:DifferentConditions}
	
	\begin{figure*}
		\includegraphics[width=0.99\textwidth]{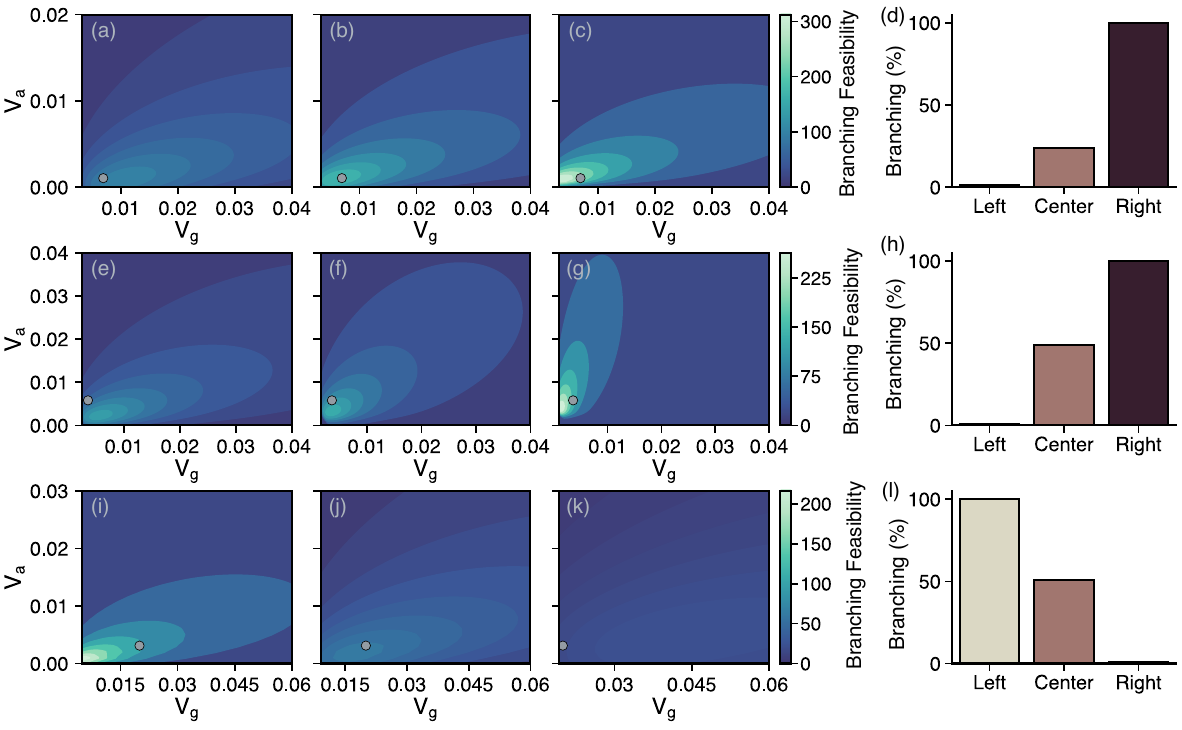}
		\caption{\textbf{Branching feasibility and the percentage of realizations that undergo branching for different experimental conditions.} a-c) Branching-feasibility plots for different values of glucose concentration in the environment supplying the chemostat: $G_0=260$ (a), $G_0=694$ (b) and $G_0=3469$ (c), all of them in $\mu$mol/L. d) Number of realizations that undergo branching at the point depicted in a-c) for the same stochastic trade-off with $(a_{tr}, b_{tr}, c_{tr}, \tau, \sigma)=(88.2,-2.5,0.014,10,6\cdot10^{-4})$ (see SM for definitions and further information). e-g) Branching-feasibility plots for different values of acetate concentration in the environment supplying the chemostat: $A_0=100$ (e), $A_0=260$ (f) and $A_0=1000$ (g), all of them in $\mu$mol/L. h) Number of realizations that undergo branching at the point depicted in e-g) for the same stochastic trade-off with $(a_{tr}, b_{tr}, c_{tr}, \tau, \sigma)=(164.3,-3.6,0.011,10,5\cdot10^{-4})$. i-k) Branching-feasibility plots for different values of the dilution rate: $D=0.0015$ (i), $D=0.006$ (j) and $D=0.015$ (k), all of them in 1/min. l) Number of realizations that undergo branching at the point depicted in i-k) for the same stochastic trade-off with $(a_{tr}, b_{tr}, c_{tr}, \tau, \sigma)=(65.5,-5.6,0.088,10,1.2\cdot10^{-4})$. Total number of realizations for the histograms is $N_r=200$. The remaining parameters specified in SM Table I where the shown $V_{g/a}$ are rescaled with respect to the table value as $V_{a/g}=\gamma V_{a/g}^\text{table}$ with $\gamma=4.5\cdot10^{10}$.}
		\label{fig:Result3}
	\end{figure*}
	As demonstrated, our model incorporating stochastic trade-offs effectively accounts for both the repeatability and variability of evolutionary branching observed in experimental data. Additionally, it provides insights into the regions in phenotypic space where branching is more likely to occur through the previously mentioned branching feasibility plots. In this section, we illustrate how these plots can be interpreted to make theoretical predictions regarding the emergence of branching under different experimental conditions.
	
	In particular, Fig.\ref{fig:Result3}a-c clearly illustrates that higher concentrations of glucose in the environment supplying the chemostat ($G_0$) generally increase the likelihood of branching. This is further verified in Fig.\ref{fig:Result3}d. Similar to Fig.\ref{fig:Result2}b-c, we use the same stochastic trade-off parameters for the three different environments whose average curve induces deterministic branching at a selected point. As evident in the histogram, the higher the glucose supply, the greater the number of realizations that undergo evolutionary branching. Moreover, the similarity in the contour pattern shape to that in Fig.\ref{fig:Result2}a is attributed to the fact that the proportion of available acetate (i.e. $\beta$ in this case) remains the same.

	On the other hand, if an acetate concentration, $A_0$, is added to the medium supplying the chemostat (Fig.\ref{fig:Result3}e-g), the pattern in the contour plot shifts, and the likelihood of branching increases in different areas of the phenotypic space. This shift occurs because the added acetate concentration increases the proportion of available acetate (i.e. $\beta+A_0/G_0$). Fig.\ref{fig:Result3}h shows the number of realizations that undergo branching for the same stochastic trade-off in the three different environments, demonstrating that environments with higher acetate supply are more likely to experience branching.

	Finally, it becomes clear from Fig.\ref{fig:Result3}i-k that the dilution rate
	is also a crucial parameter for the emergence of branching. Specifically, higher values of $D$ significantly reduce the likelihood of branching, as illustrated in Fig. \ref{fig:Result3}l.
     
   In summary, the proposed reinterpretation of the formalism developed by de Mazancourt and Dieckmann \cite{Mazancourt}, incorporating stochastic trade-offs, not only generalizes our results to encompass a wider range of trade-offs, but also enables us to compare the likelihood of observing branching under different experimental conditions. This provides a deeper understanding of the theoretical foundations of branching and may help in making more accurate predictions as well as in the design of future experiments.

	\section{Discussion} \label{sec:Discussion}
	
	Understanding the mechanisms that generate, promote and stabilize biological diversity stands as a challenging task. Microbiology experiments are particularly suited for this since the rapid adaptation/evolution of microorganisms allows one to empirically study evolution over relatively short time periods \cite{Microbiology}. Here, we focused on a particular type of experimental setup in which a well-mixed isogenic population of \emph{E. coli} is maintained on a glucose-limited chemostat over thousands of generations \cite{Helling, Rosenzweig, Treves}. After a sufficient number of generations, a number of different experimental works report a splitting of the ancestor lineage into two different ecotypes (or strains) distinguished by their carbohydrate metabolism: the first one specialises in the consumption of glucose while the second
	consumes preferentially a byproduct of glucose metabolism, i.e. acetate. This constitutes an illustrative example of the emergence of cross-feeding and the generation of complex communities/ecosystems even in the presence of few resources
	\cite{Liao, Goldford}.
	
	Previous work \cite{Doebeli-CF} showed, in the context of adaptive dynamics, that a simple eco-evolutionary model can explain in a parsimonious way such a diversification event as the result of evolutionary branching. For this one assumes a specific (hard) trade-off function between the evolving metabolic phenotypes, describing the relative preference for glucose, acetate, or a combination of both. Nevertheless, such an approach is deterministic in nature, i.e., either predicts that branching should be observed or not depending on the chosen trade-off analytical form. Therefore, it is not appropriate to explain the empirically observed stochasticity in the emergence of diversification and the variability both in branching times and in the relative weights of the emerging populations.
	
	Alternatively, in this work, we propose the idea of a stochastic trade-off that stands as an effective representation of the underlying metabolic relations that constrain the evolution of bacteria in a high-dimensional phenotypic space. This high-dimensional space includes factors beyond those directly controlling metabolic preferences.  The projection of a given mutational trajectory in such a high-dimensional manifold onto the two-dimensional phenotypic space of glucose-acetate preference does not need to be constrained to a simple curve (as illustrated in Fig. 1). 
Thus, instead of forcing the evolutionary trajectories in a lower-dimensional metabolic-preference space fixed trade-off curve, 
stochastic tradeoffs allow the populations to move in a a
wider region in such a space, introducing a new source of variability along such a space. 
In other words, the stochasticity of mutations in a high-dimensional feature space translates into variability that is not strictly constrained to a one-dimensional curve in the space of preferences, but rather to a flexible tradeoff curve.
Mathematically, the stochastic trade-off is characterized by a central curve that stands as the average of the different trajectories, the correlation time, $\tau$, that defines how fast the trajectories go back to this average curve and its amplitude, which, for fixed $\tau$, is determined by $\sigma$ and defines the width of the (fuzzy) region where populations can evolve.
	
	Thus, we have built a biologically plausible eco-evolutionary model (see Eqs.\ref{eq:Model1}-\ref{eq:Model3}), fed with experimentally measured parameters (see SM Table I) and equipped with the mentioned stochastic trade-off. We have shown that such a model is indeed able to account for the variability reported in the experiments (Fig.\ref{fig:Result1}). In particular, contrary to existing models it does reproduce: (i) the fact that branching occurs in a large proportion, but not all, of the realizations (Fig.\ref{fig:Result1}a-d) (ii) it may happen at considerably different times (Fig.\ref{fig:Result1}f) and (iii) the fraction of acetate scavengers at a certain time are variable across experiments (Fig.\ref{fig:Result1}g-h).   
	

By reinterpreting the formalism of de Mazancourt and Dieckmann \cite{Mazancourt} with the concept of a stochastic trade-off, our model generalizes the results for a broader family of trade-offs. Besides, we have developed what we term the \emph{``branching-feasibility plot''} (Fig.\ref{fig:Result2}a), which quantifies the likelihood of branching at various points in phenotypic space for a general set of stochastic trade-offs. We have verified this by plotting the number of realizations that undergo evolutionary branching at different points in phenotypic space Fig.\ref{fig:Result2}b-c. In summary, the branching-feasibility plot for our model indicates that for a wide range of stochastic trade-offs, this type of chemostat system is quite likely to undergo a diversification event, especially for low values of $V_a$. This justifies its repeated, albeit variable, emergence in actual experiments.
	
It is also noteworthy that the framework we have devised here allows for theoretical predictions regarding the likelihood of observing branching under various experimental conditions.  Indeed, in Fig.\ref{fig:Result3}, we demonstrate how different experimental parameters can influence the probability of branching and how the region in phenotypic space where branching is most likely can vary. Specifically, we have shown that increasing the amount of glucose and/or acetate supplied to the system makes evolutionary branching more likely, whereas increasing the dilution rate can make branching almost impossible. Therefore, this framework provides us with a deeper theoretical understanding of the factors controlling diversification in real experiments. Furthermore, predictions from the model can be easily tested experimentally by repeating the evolutionary experiments under different conditions and comparing the percentage of realizations that result in diversification.
	
The theoretical analyses made in this work assume, as customarily in adaptive dynamics, that evolutionary changes are perturbative in phenotypic space, i.e. phenotypes change gradually. However, this might not be the case for actual evolutionary experiments with bacteria. For example, one can imagine that mutations in certain genes could confer an initially glucose-limited bacterial population the ability to metabolize finite amounts of acetate without having to wait for many generations. Future work will therefore be devoted to implementing this possibility in the models and exploring how it can possibly change the overall picture. Finally, as mentioned in the introduction, the repeated branching of an ancestor \emph{E. coli} lineage into two distinct ecotypes characterized by their carbohydrate metabolism is also observed in serial dilution experiments in which bacteria evolve in a batch of glucose-acetate \cite{Lenski, Friesen, Spencer, Spencer-2, Tyerman, LeGac, Herron}. The framework introduced here can be easily adapted to such a scenario and further work will be devoted to it. We believe that the ideas introduced here ---in particular the idea of stochastic trade-offs--- will motivate further research into understanding the underlying principles of evolution under controlled conditions.
	
	\section{Acknowledgements} 
	This work has been supported by Grant No. PID2020-113681GB-I00 of the Spanish Ministry and Agencia Estatal de Investigación MICIU/AEI/10.13039/501100011033. It has also been supported by Grant No. PID2023-149174NB-I00 financed by MICIU/AEI/10.13039/501100011033 and EDRF funds. R.C.L. acknowledges funding from the Spanish Ministry and AEI, under Grant No. FPU19/03887 and from the Arqus Alliance of European Universities program 6.5. S.A. acknowledges the support of the NBFC to the University of Padova, funded by the Italian Ministry of University and Research, PNRR, Missione 4 Componente 2, “Dalla ricerca	all’impresa,” Investimento 1.4, Project No. CN00000033. S.S. acknowledges financial support under the National Recovery and Resilience Plan (NRRP), Mission 4, Component 2, Investment 1.1, Call for Tender No. 104 by the Italian Ministry of University and Research (MUR), funded by the European Union—NextGenerationEU—Project Title: Anchialos: Diversity, function, and resilience of Italian coastal aquifers upon global climatic changes—CUP C53D23003420001	Grant Assignment Decree No. 1015 adopted by the Italian Ministry of Ministry of University and Research (MUR). The authors warmly thank Amos Maritan for insightful discussions and for contributions to the early stages of this research.

\section{Supplementary Material}
	
\subsection{Branching feasibility plots for monomorphic populations } \label{sec:Mazancour}
We briefly review here the geometrical method developed in \cite{Mazancourt, Rueffler, Bowers} where the evolutionary outcomes of the evolution of a monomorphic population in a two-dimensional phenotypic space can be characterized without specifying a particular trade-off beforehand. This method has also been extended partially to dimorphic populations by \cite{Kisdi} and to multidimensional phenotypic spaces by  \cite{Ito}. In the following we work with a monomorphic population with two evolving phenotypes and assume the reader is familiar with the framework of adaptive dynamics which can be found in e.g. \cite{Doebeli-Book}. 

\begin{figure}[h!]
	\includegraphics[width=0.99\textwidth]{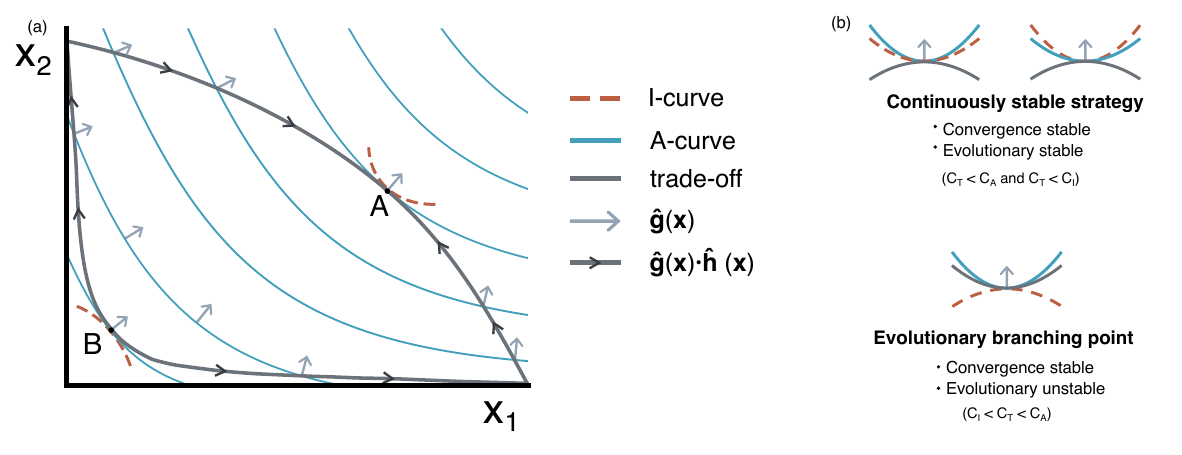}
	\caption{Geometric analysis of the possible evolutionary outcomes. a) An example of A-curves (blue lines), local I-curves (red dashed lines) and possible trade-offs (dark grey lines). The direction of the selection gradient $\textbf{g}(\textbf{x})$ (light grey arrows) and its projection onto the trade-off curves $\hat{\textbf{g}}(\textbf{x})\cdot\hat{\textbf{h}}(\textbf{x})$ (dark grey arrows) are shown to indicate the direction of selection. Singular points where the trade-off is perpendicular to the A-curve are shown in A and B. As can be seen by the direction of the arrows inside the trade-off, the former (A) is convergence stable, whereas the latter (B) is not. Besides, A is evolutionary stable since the I-curve is over the trade-off at that point whilst B is not because the I-curve is under the trade-off. b) Geometric signatures of two possible evolutionary outcomes.}
	\label{fig:Mazancour}
\end{figure}

Under frequency-dependent selection, two independent properties have to be considered: (i) evolutionary stability, i.e. strategies that cannot be invaded once established and (ii) convergence stability, i.e. strategies toward which directional evolution will converge through small evolutionary steps. Thus, four possible evolutionary outcomes exist depending on whether convergence or evolutionary stabilities are achieved or not, but we are mostly interested in two: \emph{continuously stable strategies} that share both convergence and evolutionary stability and \emph{evolutionary branching points} that have convergence but not evolutionary stability (Fig.\ref{fig:Mazancour}b). Conventional analyses of evolutionary models require ad hoc assumptions about the shape of the trade-off curves. This method, however, offers a geometric representation of evolutionary and convergence stabilities, thus enabling the visualization of how evolutionary outcomes depend on arbitrary trade-offs in an simple way. Therefore, with this method one can easily identify the possible evolutionary outcomes that general trade-offs can induce and determine the effect on them of ecological parameters independently of the specific trade-off curve.

Populations are represented in phenotypic space as points whose location correspond to the resident phenotype.  Evolution and selection change the population's phenotype over time making the point to describe a trajectory in phenotypic space (evolutionary trajectory). In the absence of frequency-dependent selection, a single fitness value can be assigned to each phenotype,  resulting  in  a fixed fitness  landscape  \cite{Lande};  a population  then just climbs up its fitness landscape until it reaches a maximum. By contrast, under frequency-dependence, where a fixed fitness landscape is absent, one relies instead on the notion of invasion fitness $f(\textbf{x}', \textbf{x})$, which is simply the per capita growth rate of the mutant's phenotype, $\textbf{x}'$ in the environment determined by the resident phenotype $\textbf{x}$. If $f(\textbf{x}', \textbf{x})$ is positive $\textbf{x}'$ can initially invade into a population dominated by $\textbf{x}$; otherwise, it cannot. Directional evolution is thus inferred from local selection gradients,  $\textbf{g}(\textbf{x})=(g_1(\textbf{x}), g_2(\textbf{x}))=\nabla_{\textbf{x}'} f(\textbf{x}', \textbf{x})|_{\textbf{x}'=\textbf{x}}$,  which describe the most-favoured direction  by  selection  around  a  population’s  resident  phenotype. 

Indeed, the phenotype $\textbf{x}$ evolves according to the canonical equation,  i.e. it attempts to climb the local selection gradient $d\textbf{x}/dt\propto\textbf{g}(\textbf{x})$ \cite{Dieckmann96}. Nevertheless, the presence of a trade-off between the evolving phenotypes, that can be written in general as $H(\textbf{V})=0$, constrains the evolution to the projection of the gradient onto the trade-off curve \cite{Ito}:
\begin{equation}\label{eq:canonical2}
	\frac{d\textbf{V}}{dt}= \mu \left(\nabla_{\textbf{V'}}f(\textbf{V'}, \textbf{V})\Bigr|_{\textbf{V'}=\textbf{V}}-\frac{\nabla H(\textbf{V})\cdot\nabla_{\textbf{V}'}f(\textbf{V'}, \textbf{V})\Bigr|_{\textbf{V'}=\textbf{V}}}{|\nabla H(\textbf{V})|^2}\nabla H(\textbf{V})\right)
\end{equation}
where $\mu$ is the evolutionary rate coefficient, that in the most general case might depend on  $\textbf{V}$ and $N$,  and $\nabla_{\textbf{V'}}=[\frac{\partial}{\partial_{V_1'}}, \ldots, \frac{\partial}{\partial_{V_P'}}]$,  where $P$ is the dimension of the phenotype space.  

In the two-dimensional case under consideration,  one has that $\textbf{x}=(x_1, x_2)$ and $x_2=h(x_1)$ where $h$ is an arbitrary trade-off function. If evolution is constrained to follow a hard trade-off curve, it  reaches an endpoint (singular point) if at some point the selection gradient is perpendicular to the tangent vector of the trade-off function, $\textbf{h}$, i.e. $\textbf{g}\cdot\textbf{h}=0$ (see points A and B in Fig. \ref{fig:Mazancour}a). Therefore, one can construct curves that, at any point in phenotypic space, are orthogonal to the local selection gradient and hence to all resultant evolutionary trajectories. These curves (blue lines in \ref{fig:Mazancour}a), called A-boundaries, are then independent from the trade-off and depend exclusively on the characteristics of the model. Is only when we are to find the specific fixed point(s) of the evolutionary dynamics that we have to specify the trade-off function. Singular points can then be seen as those in which the slopes of the A boundary, $S_A$, and the trade-off function, $S_T$, are equal or equivalently when both curves are tangent to each other. Besides, A boundaries help to distinguish in a simple geometrical way whether the fixed point is convergence stable, and as such no neighbouring phenotypes can be reached by directional evolution, (A point in Fig.\ref{fig:Mazancour}a) or unstable, where neighbouring phenotypes can indeed be reached by evolution, (B point in Fig.\ref{fig:Mazancour}a).

To asses evolutionary stability, a second type of curve, the so called local I-boundary (red dashed lines in Fig.\ref{fig:Mazancour}a), is needed. For any point in phenotypic space, $\textbf{x}$, the local I curve separates regions of mutant phenotypes $\textbf{x}'$ that can invade into a population situated at $\textbf{x}$, $f(\textbf{x}', \textbf{x})>0$, from those that cannot, $f(\textbf{x}', \textbf{x})<0$. The local I boundary is then defined as all $\textbf{x}'$ with $f(\textbf{x}', \textbf{x})=0$. Notice that at $\textbf{x}$ both A and I boundaries have the same slope, $S_A=S_I$. Indeed, at $\textbf{x}$, the slope of the A boundary, $S_A=-g_1(\textbf{x})/g_2(\textbf{x})$, can be obtained from its tangent vector, $\textbf{t}_A(\textbf{x})=\pm(g_2(\textbf{x}), -g_1(\textbf{x}))$, which is nothing but the perpendicular to the selection gradient, $\textbf{g}(\textbf{x})$. On the other hand, the slope of the local I boundary at $\textbf{x}'=\textbf{x}$ can be obtained as the slope of the implicit curve, $f_{\textbf{x}}(\textbf{x}')\equiv f(\textbf{x}', \textbf{x})=0$, which is nothing but $S_I=-\partial_{x_1'}f(\textbf{x}', \textbf{x})/\partial_{x_2'}f(\textbf{x}', \textbf{x})|_{\textbf{x}'=\textbf{x}}\equiv -g_1(\textbf{x})/g_2(\textbf{x})$.

At a singular point a resident population lying on a trade-off curve experiences disruptive selection (evolutionary instability) if its invader set (i.e. $\textbf{x}' \mid f(\textbf{x}', \textbf{x})>0$) includes the surrounding trade-off curve (B in Fig.\ref{fig:Mazancour}a). By contrast, if  the  trade-off  curve  does  not  fall  into  the  invader  set, selection on the singular point is stabilizing (A in Fig.\ref{fig:Mazancour}a).

Thus, the relative local curvatures of A-boundaries, I-boundaries, and trade-off curves at singular points determine the different possible evolutionary outcomes. In order to compare the curvatures, a convention is needed. Since, by definition, the local selection gradient at a singular point is perpendicular to the local I-boundary, A-boundary, and trade-off  curve: we define positive curvature when looking  along  the  local  selection  gradient  the  curve  is convex, otherwise the curvature is negative. Denoting the curvatures of the A, I boundaries and the trade-off by $C_A$, $C_I$ and $C_T$ respectively we can then conclude that a singular phenotype is convergence stable  (unstable)  if $C_T<C_A$ ($C_T>C_A$)  and  locally  evolutionary  stable  (unstable)  if $C_T<C_I$ ($C_T>C_I$). Thus, the continuously convergence strategy and evolutionary branching points can be easily distinguished (see Fig.\ref{fig:Mazancour}b). Analytical details corresponding to these geometric insights are provided in \cite{Mazancourt}.

In is especially interesting to discriminate the conditions for which an initially monomorphic population can become polymorphic. For this to occur, evolution must converge to an evolutionary branching point which, as we have shown, requires $C_I<C_T<C_A$. One can therefore visualize where in phenotypic space evolutionary branching is more independent from the trade-off curve by plotting at each point the difference $C_A-C_I$ (see Fig. 3a in the main text). 

\subsection{Stochastic trade-offs} \label{sec:SoftTrade}
To define the stochastic trade-off we first need to choose its average curve. We want the stochastic trade-off to give rise to evolutionary branching, therefore we shall select the average curve with the right properties to produce it. For this, we only need to define its curvature and slope at a certain point, which is the evolutionary branching point in the deterministic case (see Sec. \ref{sec:Mazancour}). Therefore, any curve that has a non trivial second order Taylor expansion at that point could be of use. To make it as simple as possible but without loss of generality, we use a second order polynomial:
\begin{equation}\label{eq:SoftTrade-off}
	V_a^{tr}(V_g)=a_{tr} V_g^2 + b_{tr} V_g + c_{tr}
\end{equation}
where the coefficients are defined such that at a certain point the polynomial has a certain curvature and slope. In particular at that point (see Sec. \ref{sec:Mazancour}):
\begin{itemize}
	\item The curvature $C_T$ has to be in the range $(C_I, C_A)$ where $C_I$ and $C_A$ are the curvatures of the I and A-curves (see Sec. \ref{sec:Mazancour}). Due to the specific form of our invasion fitness, in our case $C_I=0$ always, so $C_T=\delta C_A$ where $\delta\in(0,1)$.
	\item The slope $S_T$ must be equal to that of the A-curve $S_A$, so $S_T=S_A$.
	\item Trivially, the curve must contain the point.
\end{itemize}

Thus, denoting the selected point as $(V^s_g, V^s_a)$, the coefficients become: 

\begin{eqnarray}
	a_{tr} &=& \frac{1}{2} \delta C_A(1+S_A^2)^{\frac{3}{2}} \\
	b_{tr} &=& S_A - 2 a_{tr} V^s_g \\
	c_{tr} &=& V^s_a - b_{tr}V^s_g - a_{tr}{V^s_g}^2
\end{eqnarray}

Let us remark that these coefficients define not a single valid curve but a family of them since $\delta\in(0,1)$ is not fixed. Furthermore, as already stated before any other more complex curve with a local Taylor expansion similar to Eq.(\ref{eq:SoftTrade-off}) would also be suitable. Thus, our results can be extrapolated for a general set of average trade-off curves. However, let us note that when changing the curve, one must pay attention to avoid a curve with additional singular points (i.e. those at which evolution ends, see Sec. \ref{sec:Mazancour}) that come before in the evolutionary trajectory. These singular points can be detected by plotting the so-called pairwise invasibility plots (further details in \cite{Geritz2, PIP}). 

Finally, to fully determine the stochastic trade-off, we have to establish how the evolutionary trajectories move in the direction perpendicular to the average trade-off curve. For this, we define the distance to the average trade-off curve at each evolutionary time-step as a temporally correlated function of the distance in the previous time. Essentially, is a time-discrete Orstein-Uhlenbeck process with $\Delta t=1$ and zero average. Thus we have:

\begin{equation}\label{eq:stotrade-off}
	d_{t+1}=(1-\frac{1}{\tau})d_{t}+\sqrt{\frac{1}{\tau}}\xi
\end{equation}
where $\tau$ is the correlation time of the Ornstein-Uhlenbeck process that determine how fast the process return to the average curve (i.e to $d=0$) in the absence of the noise and $\xi$ is a random Gaussian variable of zero average and standard deviation $\sigma$ which, for fixed $\tau$, determines the amplitude (in phenotypic space) of the stochastic trade-off.

\subsection{Evolutionary algorithm} \label{sec:SimulationProcedure}

Here we follow a slightly modified version of the evolutionary algorithm employed in \cite{Caetano}. To simulate the evolution of the whole population, one needs first to extend the system of ecological dynamics Eqs.(1-3) of the main text for an arbitrary number M of sub-populations:
\begin{eqnarray}
	\frac{dN_j}{dt}&=&\gamma\left(V^j_{g}\frac{G}{K_{g}+G}+V^j_{ a}\frac{A}{K_{a}+A}\frac{C_{g}}{C_{g}+G}\right)\frac{C_{a}}{C_{a}+A}N_j-DN_j \label{eq:Model1Changed} \\
	\frac{dG}{dt}&=&D(G_0-G)-\sum_{j}V^j_{g}\frac{G}{K_{g}+G}N_j \label{eq:Model2Changed} \\
	\frac{dA}{dt}&=&\beta\sum_{j} V^j_{g}\frac{G}{K_{g}+G}N_j-\sum_{j}V^j_a\frac{A}{K_{a}+A}\frac{C_{g}}{C_{g}+G}N_j-DA 
	\label{eq:Model3Changed}
\end{eqnarray}
where $j=1, \ldots, M$ is the sub-population index. Starting from a single monomorphic population, so that $M=1$,
one  integrates the previous of equations until a steady state is reached. After that, if any of the existing sub-populations falls below a small extinction threshold density (arbitrarily fixed to $n_{min}=10^{-6}$), it is  removed from the system.
Then, a new ``mutant'' subtype is split from one of the resident sub-types, characterized by  $(V_g, V_a)$ that is randomly chosen with probability proportional to its total growth rate in the steady state. 

\begin{figure}[h!]
	\includegraphics[width=0.99\textwidth]{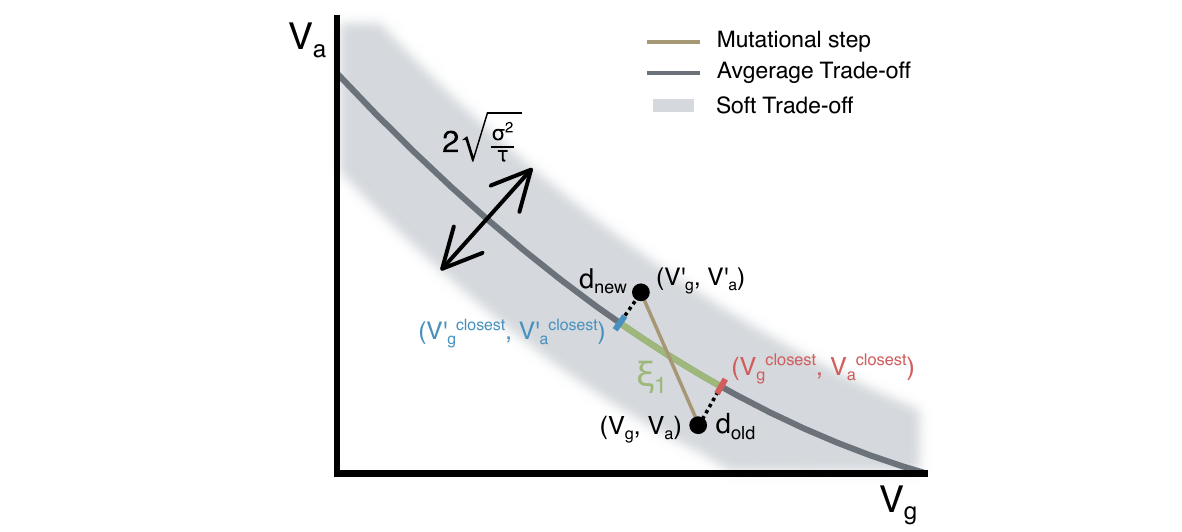}
	\caption{Representation of one mutational step of the evolutionary algorithm. In colours are represented the auxiliary variables needed to get the mutational step from the resident population at $(V_g, V_a)$ to the mutant at $(V'_g,V'_a)$. The stochastic trade-off (grey fuzzy region) and its average (dark grey curve) are also depicted with the magnitude of its amplitude $2\sqrt{\sigma^2/\tau}$.}
	\label{fig:EvolutionaryAlgorithm}
\end{figure}

To obtain the mutant phenotype $(V'_g, V'_a)$ we have to distinguish which kind of trade-off is being used. (i) In the case of the stochastic trade-off, we first determine the distance $d_{old}$ between the point $(V_g, V_a)$ and its closest point in the average trade-off curve, $(V^{closest}_g, V^{closest}_a)$ (see Fig.\ref{fig:EvolutionaryAlgorithm}). Then, we add a random mutation, $\xi_1$, selected from an uniform distribution in the range $[-m, +m]$ with $m=10^{-4}$, to $(V^{closest}_g, V^{closest}_a)$ and obtain the point that is at arc-length distance $\xi_1$ in the average curve from it (see Fig.\ref{fig:EvolutionaryAlgorithm}). Thus, one arrives to a point in the curve that we denote $(V'^{closest}_g, V'^{closest}_a)$. Finally, we obtain the mutant phenotype by using the temporarily-correlated noise (stochastic trade-off), as the point that is at distance $d_{new}=(1-\frac{1}{\tau})d_{old}+\sqrt{\frac{1}{\tau}}\xi_2$ from $(V'^{closest}_g, V'^{closest}_a)$ where $\xi_2$ is selected from a zero-averaged Gaussian distribution with standard deviation $\sigma$. (ii) On the other hand, in the case of a hard trade-off we determine the mutant phenotype as $V'_g=V_g+\xi_1$ and $V'_a=h(V'_g)$ with $\xi_1$ is defined as above and $h$ being the hard trade-off function. 

Once the mutant is obtained, its population is set to be the $10\%$ of the ancestral one, which in turn gets reduced by $10\%$. Although this fraction may seem rather large, the results do not change significantly as long as we allow sufficient time for the ecological dynamics to stabilize. Moreover, in order to reduce computational complexity, once each $\Delta t_{merge}=100$ we discretize the phenotypic space in squares of width $m$ and merge the subpopulations that fall in the same bin by summing their populations and averaging its $V_g$ and $V_a$ values.  

\subsection{Branching time determination}\label{sec:BranchingDetermination}

To determine branching times, at each evolutionary time, we pick from the $M$ subpopulations with phenotypes $\{(V^1_g,V^1_a), (V^2_g, V^2_a), \ldots ,(V^M_g, V^M_a)\}$ the subpopulation with the maximum $V^\text{max}_g$ and create a set initially containing only such value. Likewise with the minimum $V^\text{min}_g$. Then, we add to the set of the maximum (minimum) those subpopulations with $V_g$ at arc-length distance inside the curve below $d=6\cdot10^{-4}$ from the maximum (minimum). In this way, we fix a criterion to declare that branching has occurred if the two groups are disjoint and take the evolutionary time when this happens as the branching time. Besides, we denote the group with higher $V_g$ as the glucose specialist and the other ---that have a lower $V_g$ and in turn a higher $V_a$--- as the acetate scavengers.

\subsection{Evolutionary branching with a stochastic trade-off}\label{sec:BranchingSoftTradeoff}

In order to demonstrate that the evolutionary-branching feasibility plot (Fig. 3a in the main text) does indeed shows where in phenotypic space is more likely to have branching, we showed the percentage of realizations that end up causing branching at the specified points in phenotypic space (Fig. 3b in the main text). Here we explain how we have obtained Fig. 3b in the main text: (i) we choose an average curve of the stochastic trade-off that has the right properties to deterministically (in absence of the stochastic trade-off amplitude, that is, $\sigma=0$) originate evolutionary branching at a certain point, $\textbf{V}^s=(V^s_g,V^s_a)$  (for example the points marked with a cross in Fig. 3a in the main text). This can be easily done as detailed in Methods Sec. \ref{sec:SoftTrade}. (ii) Then, once we have the stochastic trade-off defined with a certain $\sigma$ and $\alpha$, we initiate the evolutionary algorithm with a monomorphic population whose phenotype, is equal to that of the selected point, $\textbf{V}^\text{init}=\textbf{V}^s$. We do it in this way to be able to compare the branching times for the different point that we select. (iii) At $t^\text{max}_\text{evo}$ (which can vary as specified in the label of Fig. 3 in the main text) we determine whether two phenotypically distinguishable groups of subpopulations emerged as described in Methods Sec.\ref{sec:BranchingDetermination}. (iv) We repeat this process for all the different realizations and stochastic trade-off amplitudes $\sigma$.

\subsection{Parameter values}
\begin{table}[h!]
	\centering
	\begin{tabular}{ p{5cm} p{1cm} p{4.3cm} p{1cm}  }
		\multicolumn{4}{c}{\textbf{Model parameter values}} \\
		\hline
		Parameter description & Parameter notation & Parameter value & Reference \\
		\hline
		Order of magnitude for maximal glucose/acetate uptake rate* & $V^\text{table}_{g/a}$ & $\sim$\SI{5e-13}{\umol\per\minute\per\cell} & \cite{Rosenzweig}\\
		Half saturation constant of glucose uptake & $K_{g}$ & \SI{10}{\umol\per\liter} & \cite{Rosenzweig}\\
		Half saturation constant of acetate uptake & $K_{a}$ & \SI{200}{\umol\per\liter} & \cite{Rosenzweig}\\
		Acetate inhibition constant & $C_a$ & \SI{e5}{\umol\per\liter} & \cite{Gudelj}\\
		Glucose repression constant & $C_g$ & \SI{4e4}{\umol\per\liter} & \cite{Gudelj}\\
		Chemostat dilution rate & $D$ & \SI{0.003}{\per\minute} & \cite{Rosenzweig}\\
		Glucose concentration in ambient supplying the chemostat & $G_0$ & \SI{346.9}{\umol\per\liter} & \cite{Rosenzweig}\\
		Growth proportionality constant & $\gamma$ & \SI{4.5e10}{\cell\per\umol} & \cite{Gudelj} \\           
		\hline
	\end{tabular}
	\caption{Experimental values of the model parameters. *Since $V_{g/a}$ are the evolving parameters, we only need its order of magnitude. Besides, in order to simplify the values, for all our results we show a rescaled version of $V_{g/a}$, that is, $V_{g/a}=\gamma V_{g/a}^\text{table}$.}
	\label{table:ExValues}
\end{table}



\end{document}